\documentclass[conference]{IEEEtran}
\IEEEoverridecommandlockouts
\usepackage{cite}
\usepackage{amsmath,amssymb,amsfonts}
\usepackage{algorithmic}
\usepackage{graphicx}
\usepackage{textcomp}
\usepackage{xcolor}
\def\BibTeX{{\rm B\kern-.05em{\sc i\kern-.025em b}\kern-.08em
    T\kern-.1667em\lower.7ex\hbox{E}\kern-.125emX}}

\usepackage[shortlabels]{enumitem}
\usepackage{amsthm}
\usepackage[position=top]{subfig}
\usepackage{subfloat}
\usepackage{xcolor}
\usepackage{multicol}
\usepackage{multirow}
\usepackage{hhline}
\usepackage{soul}
\usepackage{tabularx}
\usepackage{xspace}
\theoremstyle{definition}
\newtheorem{definition}{Definition}[subsection]

\usepackage{tikz}
\usetikzlibrary{arrows.meta,
                calc, chains,
                quotes,
                positioning,
                shapes.geometric}
                
\usepackage{comment}    
\usepackage{url}
\usepackage{balance}

\newcommand{\inlinedComment}[2]{\textcolor{#1}{\small\textbf{#2}}} 
\newcommand{\eric}[1]{\inlinedComment{blue}{Eric says: #1}}

\newcommand{\ourapp}{\textsc{MANDO}\xspace}

\begin{document}

\title{\huge MANDO: Multi-Level Heterogeneous Graph Embeddings for Fine-Grained Detection of Smart Contract Vulnerabilities}


\author{
\IEEEauthorblockN{Hoang H. Nguyen$^\dagger$, Nhat-Minh Nguyen$^\ddagger$, Chunyao Xie$^\dagger$, Zahra Ahmadi$^\dagger$, Daniel Kudendo$^\dagger$,\\Thanh-Nam Doan$^\S$, Lingxiao Jiang$^\ddagger$
\thanks{\textbf{Acknowledgment.} This work was supported by the European Union’s Horizon 2020 research and innovation program under grant agreement No. 833635 (project ROXANNE: Real-time network, text, and speaker analytics for combating organized crime, 2019-2022)
and by the Singapore Ministry of Education (MOE) Academic Research Fund (AcRF) Tier 1 grant.}}
\IEEEauthorblockA{\textit{$^\dagger$L3S Research Center, Leibniz Universität Hannover, Hannover, Germany} \\
\textit{$^\ddagger$Singapore Management University, Singapore}\\
\textit{$^\S$Independent Researcher, Atlanta, Georgia, USA} \\
\{ehoang,xie,ahmadi,kudendo\}@l3s.de, \{nmnguyen,lxjiang\}@smu.edu.sg, me@tndoan.com
}
}

\maketitle

\begin{abstract}
Learning heterogeneous graphs consisting of different types of nodes and edges enhances the results of homogeneous graph techniques.
An interesting example of such graphs is control-flow graphs representing possible software code execution flows.
As such graphs represent more semantic information of code, developing techniques and tools for such graphs can be highly beneficial for detecting vulnerabilities in software for its reliability.
However, existing heterogeneous graph techniques are still insufficient in handling complex graphs where the number of different types of nodes and edges is large and variable. 
This paper concentrates on the Ethereum smart contracts as a sample of software codes represented by \emph{heterogeneous contract graphs} built upon both control-flow graphs and call graphs containing different types of nodes and links.
We propose \ourapp, a new heterogeneous graph representation 
to learn such heterogeneous contract graphs' structures.
\ourapp extracts customized metapaths, which compose relational connections between different types of nodes and their neighbors. Moreover, it develops a multi-metapath heterogeneous graph attention network to learn multi-level embeddings of different types of nodes and their metapaths in the heterogeneous contract graphs, which can capture the code semantics of smart contracts more accurately and facilitate both fine-grained line-level and coarse-grained
contract-level vulnerability detection.
Our extensive evaluation of large smart contract datasets shows that \ourapp improves the vulnerability detection results of other techniques at the coarse-grained contract level.
More importantly, it is the first learning-based approach capable of identifying vulnerabilities at the fine-grained \textit{line-level}, and significantly improves the traditional code analysis-based vulnerability detection approaches by 11.35\% to 70.81\% in terms of F1-score.

\end{abstract}

\begin{IEEEkeywords}
heterogeneous graphs, graph embedding, graph neural networks, vulnerability detection, smart contracts, Ethereum blockchain
\end{IEEEkeywords}

\begin{figure*}[!t]
\centering
\includegraphics[scale = 0.21]{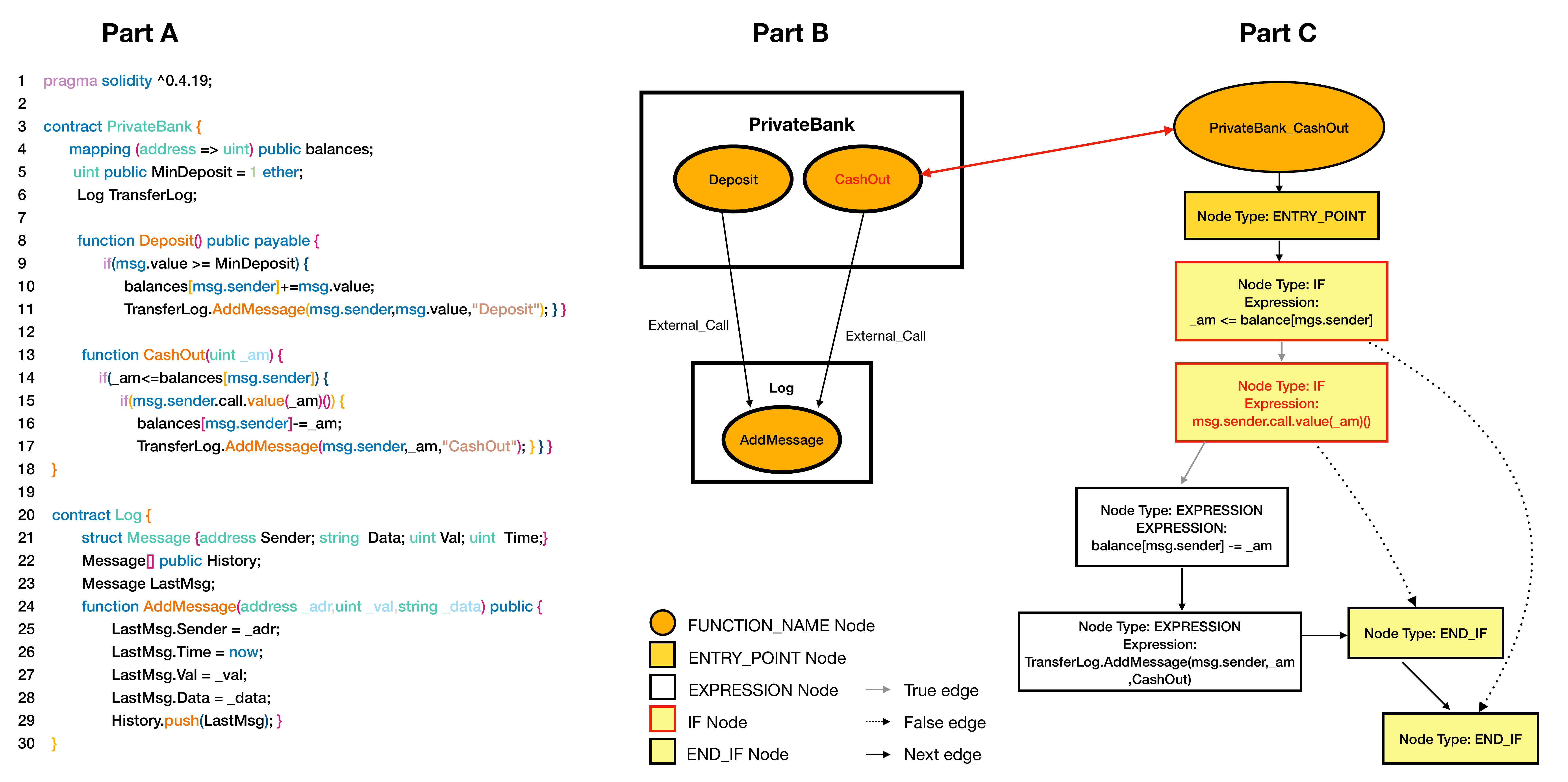}
\caption{\small A sample Ethereum smart contract code snippet (Part A), its corresponding heterogeneous call graph (CG) (Part B), and a sample heterogeneous control-flow graph (CFG) for the function \texttt{CashOut} in the contract \texttt{PrivateBank} (Part C).
Line 15 in Part A is the root cause of a Reentrancy bug;
the nodes in CG and CFG containing the Reentrancy bug are highlighted with red text}.
\label{fig:code-snippet}
\end{figure*}

\section{Introduction} \label{sec:intro}

Graph learning has been an active research area for a long time. 
Learning \emph{heterogeneous} graphs that consist of nodes and edges of different types has recently attracted extensive attention since such graphs contain richer information from the application domains than homogeneous graphs and, therefore, can achieve better learning results~\cite{HAN}. However, when it comes to complex heterogeneous graphs, where the graph structures have particular properties and the number of node types and edge types can be arbitrarily large and changing, it is still unclear if existing techniques can handle them well. Examples of such graphs can be found in control-flow graphs or call graphs representing possible software code execution flows and call relations. A control-flow graph depicts all possible sequences of statements or lines of code that might be traversed in one function during program executions. In contrast, a call graph represents every possible call relation among functions in a program.


This paper aims to develop a new approach for learning such complex and dynamic heterogeneous graphs and apply them to address critical software quality assurance problems, such as detecting vulnerabilities in software code that can be represented as control-flow graphs and call graphs.
Expressly, we represent software code as a combination of heterogeneous graphs of multiple granularity levels that capture the control-flow and call relations in code. 
Then, we extract specially defined \emph{metapaths}
for such graphs that acquire relations between different types of nodes and their neighbors, 
and fuse various kinds of graph neural networks together to learn both of the node-level and graph-level embeddings.
Further, we use the embeddings to train networks to recognize graphs or nodes that may contain vulnerabilities and thus identify the vulnerable code functions or lines. 
Last but not least, we apply our approach to the Ethereum smart contracts written in the Solidity programming language. 
We choose smart contracts from distributed blockchains \cite{wood2014ethereum} 
as they become increasingly popular in various domains that involve payments and contracts. Different techniques are essential to detect their potential bugs and ensure correct executions of the payments and contracts. 
In short, our approach enables novel 
\underline{m}ulti-level gr\underline{a}ph embeddi\underline{n}gs for fine-grained \underline{d}etection of smart c\underline{o}ntract vulnerabilities, 
and thus we name it as \ourapp. 
\ourapp is novel in its graph neural network structure that fuses topological GNN and node-level attentions with heterogeneous GNN to generate both node-level and graph-level embeddings that can capture structural information of graphs more accurately. It is also novel in enabling both node-level and graph-level classifications to detect fine-grained line-level vulnerabilities in smart contract source code in addition to coarse-grained contract-level vulnerabilities.

For our empirical evaluation, we have curated a mixed dataset containing 493 Solidity vulnerable contracts from multiple data sources from previous studies.
There are seven types of vulnerabilities in the dataset; each has between 50 to 80 instances. 
Our evaluation results show that \ourapp achieves a heightened F1-score from 81.98\% to 90.51\% for detecting the vulnerabilities at the fine-grained line-level, while previous deep learning and embedding-based techniques can only detect the vulnerabilities at the contract file/function level.
We also show that, compared to a few different graph embedding models (such as node2vec~\cite{node2vec}, LINE~\cite{tang2015line}, GCN~\cite{GCN}, and metapath2vec~\cite{metapath2vec})
and traditional program analysis techniques (such as Securify~\cite{tsankov2018securify}, Mythril~\cite{Mueller2018}, Slither~\cite{Slither}, Manticore~\cite{mossberg2019manticore}, Smartcheck~\cite{tikhomirov2018smartcheck}, and Oyente~\cite{Oyente}) that can detect vulnerabilities at the line level, our method improves their F1-score by 11.35\% to 70.81\% for various bug types. 

\begin{figure*}[!htb]
\centering
\includegraphics[scale = 0.41]{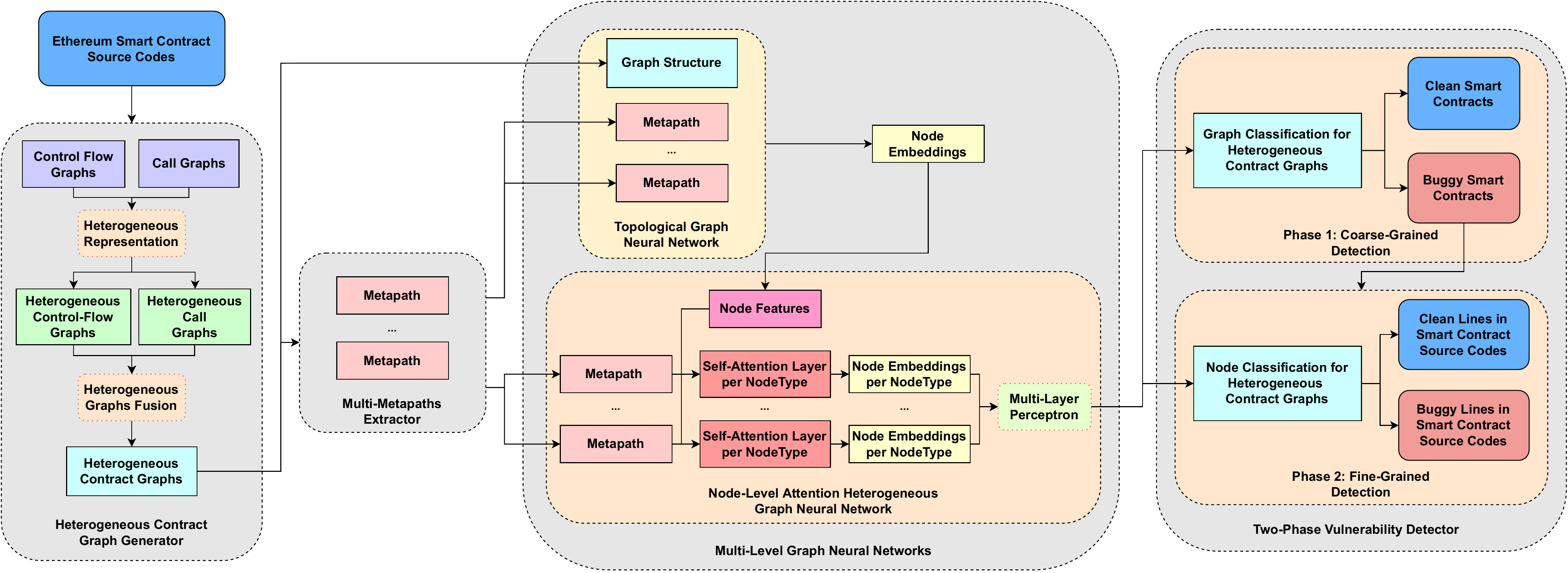}
\caption{\small Overview of the \ourapp framework.}
\label{fig:overview}
\end{figure*}

To summarize, our main contributions are as follows:
\begin{itemize}[leftmargin=1em,nosep]
\item We propose a new technique for representing Ethereum smart contracts written in Solidity as {\em heterogeneous contract graphs} that combines heterogeneous control-flow graphs (CFGs) and call graphs (CGs) of multiple levels of granularity. This new technique allows us to represent the semantic relation of node and edge types that the previous approaches could not capture with only using the homogeneous forms of these CFGs and CGs separately. 
\item We propose a novel architecture for the Heterogeneous Graph Neural Network using Node-Level Attention (Figure~\ref{fig:overview} and \ref{fig:mando-hgnn}), which fits our customized metapaths, to build 
embeddings of multiple granularity levels for heterogeneous contract graphs.
\item We employ the multi-level embeddings of heterogeneous graphs and labeled instances of vulnerable smart contracts to detect new vulnerabilities accurately at the line-level and contract-level, achieving better results than prior state-of-the-art bug detection techniques for smart contracts.
\item We also publicize the dataset and our graph embedding models for the research community\footnote{\url{https://github.com/MANDO-Project/ge-sc}}.
\end{itemize}

The rest of the paper is organized as follows:
Section~\ref{sec:problem} defines our main research problem and objective with a motivating example.
Section~\ref{sec:approach} describes the detailed structure of \ourapp.
Section~\ref{sec:eval} presents the experimental settings and results to show the effectiveness of our method. 
Section~\ref{sec:related} reviews the related studies.
Section~\ref{sec:con} concludes our paper with some discussions on its limitations and future outlook.

\section{Motivation and Problem Definition}
\label{sec:problem}
%
{\bf Motivating Example}: 
Figure~\ref{fig:code-snippet} (Part A) shows a sample code snippet of a smart contract written in Solidity.
Part B shows the corresponding call graph (CG) of the contract.
Part C shows a partial sample control-flow graph (CFG) for the \texttt{CashOut} function
containing a vulnerability whose root cause is at Line 15 as {\tt msg.sender.call} can repeatedly trigger calls to {\tt CashOut} before {\tt balances} is deducted at Line 16, which means {\tt msg.sender} can receive more values than what is specified by {\tt \_am}.
In order to catch this so-called {\it reentrance} vulnerability, the control-flow and call relations among {\tt msg.sender}, {\tt balances}, and {\tt \_am} should be considered.
We aim to automatically capture such vulnerabilities' properties via our new graph embedding techniques. 

{\bf Problem Statement}:
Our high-level problem is to develop more effective heterogeneous graph learning techniques, and use them to detect fine-grained line-level software vulnerabilities and their types. 
More specifically, our objective for smart contracts written in Solidity based on our unique graph representation and embedding techniques is to: 
(1) Represent it as a \emph{heterogeneous contract graph} that combines its control-flow graph and call graph like the example in Figure~\ref{fig:code-snippet};
(2) Learn the embeddings of the graphs and the nodes at multiple levels of granularity to capture the syntactical and semantic information of smart contract code;
(3) 
Accurately identify the nodes that contain certain types of vulnerabilities and locate them in the contract code.

{\bf Usage Scenarios}:
Such
accurate vulnerability detection can be useful for smart contract quality assurance under various situations. For example:
\begin{itemize}[nosep,leftmargin=1em]
\item During the contract development in an Integrated Development Environment (IDE), it
can help to identify early if the contract contains any vulnerability of known types.
\item When a developer is reusing a contract from a third party, the vulnerability detection can check if it contains any known vulnerabilities and warns the developer about potential risks in reusing the contract directly. 

\item Whenever a new type of vulnerability is discovered, we may want to audit all existing contracts again to check if they contain the new type of vulnerability. The vulnerability detection can then be easily applied to all the contracts on a large scale for this purpose.
\end{itemize}
We believe that \ourapp can be adapted to other software as long as their control-flow and call graphs can be constructed and there are vulnerability datasets available for training. 

\section{The \ourapp Approach}
\label{sec:approach}
\subsection{Overview}
This section gives an overview of our proposed approach consisting of four main components presented in the four grey boxes in Figure~\ref{fig:overview} and describe each component in the following subsections. 
The input of our approach is the source code of one or many Ethereum smart contract source files written in Solidity.
The output is the bug prediction and the bug line in the source code if there is one.

First, the source code is processed by the \textbf{Heterogeneous Contract Graph Generator} component and translated into two heterogeneous graphs based on call graphs and control-flow graphs corresponding to two levels of granularity: contract level and statement (line) level, respectively. 
Then,  the two heterogeneous graphs are fed into the second component: \textbf{Multi-Metapaths Extractor}. 
Based on the type of each node and the types of its associated edges, the component extracts their corresponding \emph{metapaths}.
This component is novel in the sense that it can handle dynamic numbers of node and edge types in metapaths from the automatically generated heterogeneous contract graphs. 
The third component, \textbf{Multi-Level Graph Neural Networks}, contains two steps. The first step takes metapaths or graph topology of the contract graphs from the previous component as input and generates node embeddings.
Then, in the second step, the node embeddings are used as node features and fused with metapaths using heterogeneous attention mechanisms at the node level.
\textbf{Two-Phase Vulnerability Detector}, the last component, uses the embeddings to train multi-layer perceptron (MLP) to perform either graph classification or node classification, depending on the kind of the input heterogeneous contract graphs.
In \textbf{Coarse-Grained Detection}, the heterogeneous contract graphs embeddings are used to classify graphs if their respective contract is clean or vulnerable. In \textbf{Fine-Grained Detection}, the heterogeneous contract graphs embeddings of the vulnerable contracts, classified in the first phase, are used to classify a node of a contract graph as to whether it is clean or vulnerable.  
The classified nodes can then be used to find the exact locations of the vulnerabilities in specific contracts (i.e., contract-level) and specific statements or lines of code (i.e., line-level). 
\subsection{Heterogeneous Contract Graph Generator}
\label{subsec:contract-graph-generator}
Our approach uses Slither~\cite{Slither} to traverse and analyze the source code of each Ethereum smart contract for generating the basic control-flow graphs and call graphs with homogeneous structures where nodes and edges have no types or labels. 
Then, we transform these constructed graphs into heterogeneous forms to represent the semantics of graph structures and the relation of different node and edge types:
\begin{definition}[Heterogeneous Graph] 
A heterogeneous graph is a directed graph $G = (V, E, \phi, \psi)$, consisting of a vertex set $V$ and an edge set $E$. $\phi: V \rightarrow A$ is a node-type mapping function and $\psi: E \rightarrow R$ is an edge-type mapping function.
$A$ and $R$ denote the sets of node types and edge types, and $|A|\ge 2$ and $|R| \ge 1$.
\end{definition}

\textbf{Heterogeneous Control-Flow Graphs (HCFGs).}
A control-flow graph of a function is an intermediate representation of all possible sequences of statements or lines of code that might be traversed when the function is executed, which is widely used in program analysis methods.
Recent approaches on smart contract vulnerability detection use such graph representations of code when applying graph neural networks \cite{zhuang2020smart,liu2021smart}, but they mostly normalize and convert those representations into homogeneous graphs before applying graph models.
In particular, they only keep the major nodes and eliminate some normal nodes to normalize graphs since using nodes of diverse code semantics brings difficulties in training their graph neural networks. 
Thus, these approaches tend to lose valuable information regarding the source code semantics in smart contracts.
In contrast, \ourapp focuses on retaining most of the structure and semantics of the source code through heterogeneous representations where a variety of node types and edge types are preserved, called \textit{heterogeneous control-flow graphs}. 

The set of all node types in control-flow graphs is denoted as $A_{{CF}}$. Some typical node types include ENTRY\_POINT, EXPRESSION, NEW VARIABLE, RETURN, IF, END\_IF, IF\_LOOP, and END\_LOOP. 
Additionally, diverse types of connections among nodes are used to describe statements' sequential or branching structure through edge types such as NEXT, TRUE, FALSE. The set of all edge types in control-flow graphs is $R_{{CF}}$.
Figure~\ref{fig:code-snippet} (Part C) shows
a sample heterogeneous control-flow graph generated for the \textit{CashOut} function of contract PrivateBank. A Solidity parser (e.g., Slither) produces the complete sets of $A_{{CF}}$ and  $R_{{CF}}$ based on the grammar of the Solidity language. 
$G_{CF} = \{V_{CF}, E_{CF}, \phi_{CF}, \psi_{CF} \}$ denotes an HCFG with $V_{CF}$ and $E_{CF}$ as its vertex and edge sets, respectively. Each node $i \in V_{CF}$ can be viewed as a tuple of $(i, \phi_{CF}^i)$, where $i$ is the index of node and $\phi_{CF}^i \in A_{{CF}}$ is the type of node $i$. Similarly, each edge $(i, j) \in E_{CF}$ has an edge type $\psi_{CF}^{i,j} \in R_{{CF}}$. 
Each function in a smart contract can have an HCFG generated for it, and the HCFG has an entry node corresponding to the entry point/header of the function. A smart contract may be viewed as a set of HCFGs as it may contain more than one function.

\textbf{Heterogeneous Call Graphs (HCGs).} Call graphs are an intermediate representation of invocation relations among functions from the same smart contract or different smart contracts. 
A call graph generated via static program analysis often represents every possible call relation among functions in a program. 
Our study focuses on two major types of calls in smart contracts: {\em internal calls} for function calls inside one smart contract and {\em external calls} for function calls from a contract to others, represented by the two respective edge types INTERNAL\_CALL and EXTERNAL\_CALL. 
In addition, Solidity fallback functions are important in Ethereum blockchain, executed when a function identifier does not match any of the available functions in a smart contract or if no suitable data was provided for the function call. Many vulnerabilities in Ethereum smart contracts are directly or indirectly related to such fallback functions~\cite{chen2020survey}. Therefore, we represent such fallback functions with a particular node type, called FALLBACK\_NODE, besides the typical function node type FUNCTION\_NAME.

One HCG is generated from each smart contract.
$G_{C} = \{V_C, E_C, \allowbreak \phi_C, \psi_C\}$ denotes a heterogeneous call graph with $V_C$ and $E_C$ as its node and edge sets, respectively. Each node $i$ in $V_C$ can be viewed as a tuple $(i, \phi_C^i)$ where $i$ is the index of node, $\phi_C^i \in A_C$ is the type of the node $i$ and $A_C$ is the set of all node types in $G_C$.
Similarly, each edge $(i, j) \in E_{C}$ has an associate edge type $\psi_{C}^{i,j} \in R_{{C}}$.

\textbf{Heterogeneous Contract Graphs: Fusion of Heterogeneous Call Graphs and Heterogeneous Control-Flow Graphs.}
The structures of these two graphs for a smart contract can be shared or combined into a global graph to enrich information for learning. In \ourapp, we design a core for HCGs and HCFGs fusion. 
Accordingly, the HCG edges of the smart contract act as bridges to link the discrete HCFGs of the smart contract functions into a global fused graph. Specifically, the fusion graph of the heterogeneous CG and the heterogeneous CFGs for a smart contract is denoted by $G_{Fusion} = \{V_F, E_F, \phi_F, \psi_F\}$, where $V_{F} = V_{C} \cup V_{CF}^{1} \cup ... \cup V_{CF}^{N}$ and $E_{F} = E_{C} \cup E_{CF}^{1} \cup ... \cup E_{CF}^{N}$, and $N$ is number of the  HCFGs for the contract.
Intuitively, for each and every function node $i$ in the call graph, the function control-flow graph $G^i_{CF}$ is attached to the function node $i$ at the entry node of $G^i_{CF}$, and thus the call graph is expanded with control-flow graphs to produce the heterogeneous contract graph.
For example, in Figure \ref{fig:code-snippet}, the red arrow between {\tt CashOut} in Part B and {\tt PrivateBank\_CashOut} in Part C indicates a sample fusion between CGs and CFGs.

\subsection{Multi-Metapaths Extractor}
\begin{definition}[Metapath] 
A metapath $\theta$ is a path in the form of $A_{1} \xrightarrow[]{R_{1}} A_{2} \xrightarrow[]{R_{2}} ... \xrightarrow[]{R_{l}} A_{l+1}$, which defines a composite relation $R =  R_{1} \circ R_{2} \circ ... \circ R_{l}$ between type $A_{1}$ and $A_{l+1}$, and $\circ$ denotes the composition operator on relations. Note that, the \textbf{length} of $\theta$ is the number of relations in $\theta$.
\end{definition}
The number of node type in our generated graphs is dynamic, and can reach sixteen, with three distinct connection types per node type, especially in the heterogeneous control-flow graphs. 
Pre-defining all possible metapaths with any length according to all possible node types and edge types is a challenge,
as it would lead to exponential explosion of metapaths, increased data sparsity, and reduced training accuracy. 
For example, in Figure~\ref{fig:code-snippet}, between a node of ENTRY\_POINT type and a node of EXPRESSION type, several different node types can be included, such as IF and END\_IF, and in other smart contracts, NEW\_VARIABLE, IF\_LOOP, and END\_LOOP can also be included.
Besides, the order of these node types can change dynamically, depending on the input contracts' structures.

In order to address the problem of exploding and changing metapaths, our method focuses on length-2 metapaths through reflective connections between adjacent nodes to extract multiple metapaths. For instance,  the relation between two adjacent nodes of the types ENTRY\_POINT and IF in Figure~\ref{fig:code-snippet} can be described by a length-2 metapath: $ENTRY\_POINT \xrightarrow[]{next} IF \xrightarrow[]{back} ENTRY\_POINT$. 
HCFGs are mostly tree-like, having very few of their own back-edges induced by the LOOP-related statements in the source code. This can lead to the lack of metapaths connecting many leaf-node types in the graphs. Adding the ``back'' relations helps alleviate the lack and improves the completeness of the extracted metapaths.

Previous studies \cite{HAN,sun2011pathsim} also used length-2 in their evaluation, and a length-N metapath can be decomposed into (N - 1) length-2 metapaths. 
Thus, we follow those studies by using length-2 to capture the unique semantic between each node types pair and their neighbors and leave longer metapaths for future evaluations.
Similar to the methods used in HAN~\cite{HAN}, we extract the set of length-2 metapaths of each node types pair in a smart contract.




\subsection{Multi-Level Graph Neural Networks}
\label{subsec:multi-level-gnn}
\begin{figure}[tb]
\centering
\includegraphics[scale = 0.85]{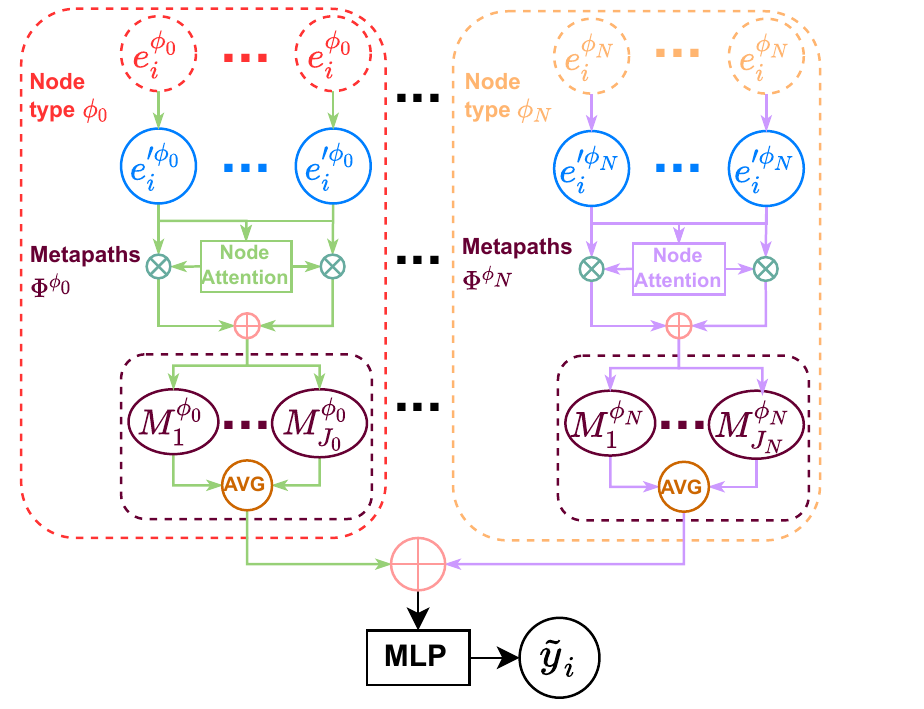}
\caption{\small Our Novel Architecture for Node-Level Attention Heterogeneous Graph Neural Network in the \ourapp Framework.}
\label{fig:mando-hgnn}
\end{figure}

\begin{table}[!t]
    \centering
    \footnotesize
    \begin{tabular}{l|l}
         Notation & Explanation \\
         \hline
         $i$                & Node $i$ \\
         $\phi_k$           & Node type $k$ \\
         $e^{\phi_k}_i$     & Node embedding of $i$ whose type is $\phi_k$\\
         $e'^{\phi_k}_i$    & Linear transformation of $e^{\phi_k}_i$ \\
         $W_{\phi_k}$       & Matrix transformation for node $i$ with type $\phi_k$\\
         $\Phi_t^{\phi_{k}}$& $t$-th metapath of node type $\phi_k$ \\
         $M_{i_t}^{\phi_k}$ & $t$-th metapath embedding of node $i$ whose node type is $\phi_k$ \\
         $M_{i}^{\phi_k}$   & Embedding of node type $\phi_k$ of node $i$\\
         $N^{\phi_k}$       & A set of metapath of node type $\phi_k$\\
         $J_k$              & Total index of node type $\phi_k$
    \end{tabular}
    \caption{\small Table of Notation}
    \label{tab:notation}
\end{table}

This component has two major building blocks: \textit{Topological Graph Neural Network} and \textit{Node-Level Attention Heterogeneous Graph Neural Network}. The former learns an input graph topology, while the latter weights the importance of the metapaths in the graph.

\subsubsection{Topological Graph Neural Network}

The main goal of this building block is to capture the graph topology. 
Each node $i$ has a node embedding $e_i$ such that $e_i$ and the embedding vector $e_j$ of the neighboring nodes $j$ of $i$ are near in the embedding space. 
Various state-of-the-art neural network techniques can be used to generate node embeddings of graphs. For a more comprehensive comparison of their effectiveness, we employ both embedding techniques for homogeneous graphs (e.g., node2vec \cite{node2vec}) and embedding techniques for heterogeneous graphs (e.g., metapath2vec \cite{metapath2vec}) in our empirical evaluation (see Section~\ref{sec:eval}).


\subsubsection{Node-Level Attention Heterogeneous Graph Neural Network}

There are two kinds of input sources for this building block: the node embeddings from the previous topological graph neural network and the metapaths from the Multi-Metapaths Extractor.

\textbf{Node-Level Attention Graph Neural Network.} Inspired by the node-level attention mechanism proposed by HAN~\cite{HAN}, we also learn to weigh the importance of every metapath and node.
However, unlike HAN, our novel approach can handle multiple dynamic customized metapaths without pre-defining the list of input metapaths (see Figure~\ref{fig:mando-hgnn} and the summary of notations in Table~\ref{tab:notation}).
The previous topological graph neural network produces a node embedding $e^{\phi_k}_i$ for each node $i$ whose type is $\phi_k$; then, we construct a corresponding weighted node feature $e'^{\phi_k}_i$ by the following linear transformation:
\begin{equation}
    e'^{\phi_k}_i = W_{\phi_k} e^{\phi_k}_i, 
\end{equation}
where $W_{\phi_k}$ is the transformation matrix associated to the type $\phi_k$ of node $i$. Each node type $\phi_k$ has a specific matrix $W_{\phi_k}$ to increase the flexibility of the transformation by projecting each type into a separated weight space.

We measure the weight of the $t$-th metapath $\Phi_t^{\phi_{k}}$  according to the node type $\phi_k$ of  $(i, j)$ pair by leveraging the self-attention mechanism \cite{transformer} between $i$ and $j$.
The weight $a^{\Phi_t^{\phi_k}}_{ij}$ is defined as follows:
\begin{equation}
    a^{\Phi_t^{\phi_k}}_{ij} = \text{softmax}_j (\textsc{att}([e'^{\phi_k}_i, e'^{\phi_k}_j] ; \Phi_t^{\phi_{k}})), 
    \label{eq:attention}
\end{equation}
where \textsc{att} is a multi-layer perceptron~\cite{DeepLearning} whose values of parameters are automatically learned through back-propagation. The input of such perceptron is the concatenation of two vectors $e'^{\phi_k}_i$ and $e'^{\phi_k}_j$. We then normalize the output of \textsc{att} into the range between 0 and 1 by all neighbors of $j$ in metapaths. 
The $t$-th metapath embedding $M_{i_t}^{\phi_k}$ of node $i$ whose node type is $\phi_k$ is a weighted sum of the node features of its neighbors with corresponding weights defined in Equation (\ref{eq:attention}). The formula is as follows:
\begin{equation}
    M_{i_t}^{\phi_k} = \sigma \left( \sum_{j\in \mathcal{N}^{\Phi_k}_i} a^{\Phi_t^{\phi_k}}_{ij} \cdot e'^{\phi_k}_j \right), 
    \label{eq:metapathEmb}
\end{equation}
where $\sigma$ is the activation function, and $\mathcal{N}^{\Phi_k}_i$ denotes the neighbors of the node $i$ according to the metapath $\Phi_k$.

To overcome the obstacle of high variance of data in heterogeneous graphs, we propose to aggregate multi-metapath embeddings with different types of nodes.
Particularly, the metapath embedding $M_{i_t}^{\phi_k}$ of each node in Equation (\ref{eq:metapathEmb}) is calculated $N$ times and then concatenated to create a final embedding ${M'}_{i_t}^{\phi_k}$ for each metapath. 



After extracting the metapath embedding, we calculate the corresponding embedding of node $i$ by averaging all metapath embedding related to $i$, noted AVG in Figure~\ref{fig:mando-hgnn}. Specifically, the embedding of node $i$ with node type $\phi_k$ is:

\begin{equation}
    M_{i}^{\phi_k} = \frac{\sum_t M_{i_t}^{\phi_k}}{|N^{\phi_k}|}, 
\end{equation}
where $N^{\phi_k}$ is a set of metapaths of the node type $\phi_k$, and the total index of the node type $\phi_k$ is equal to the size of this set i.e., $|N^{\phi_k}|$.

For fine-grained detection, we concatenate all node embedding $M_{i}^{\phi_k}$ corresponding to all node type $\phi_k$ of all node $i$ to generate a unified embedding vector for a node. We get the average of all node embeddings belonging to the graph for coarse-grained detection.

\subsubsection{Optimization for Detection}
\label{sec:optimize}
We employ the multi-layer perceptron (MLP) with a softmax activation function for the graph and node classification tasks. The input of such a layer is dependent on the type of prediction tasks. 
The loss function for the training process is cross-entropy, and the parameters of our model are learned through back-propagation.



\subsection{Two-Phase Vulnerability Detector}
\label{sec:twophasesdetector}
This component has two main phases: \textit{Coarse-Grained Detection} and \textit{Fine-Grained Detection}. The first phase classifies clean versus vulnerable smart contracts at the coarse-grained contract level; the second phase identifies the actual locations of the vulnerabilities in the smart contract source code at the fine-grained line level.
Providing line-level locations of the vulnerabilities is one of our primary contributions,
while the previous graph learning-based methods~\cite{zhuang2020smart,liu2021combining} only report vulnerabilities at the contract or function level.

\subsubsection{Phase 1: Coarse-Grained Detection}
\label{detector:phase1}
This phase classifies if a smart contract contains a vulnerability.
We use the fused heterogeneous call graphs and control-flow graphs (i.e., heterogeneous contract graphs) and their embeddings to represent each input smart contract, and train the MLP (Section \ref{sec:optimize}) to predict clean or vulnerable contracts.
As there can be many clean smart contracts, this classification assists in reducing the search space by filtering out those clean contracts and reducing noisy data before the second phase of fine-grained vulnerability detection at the line level.

\subsubsection{Phase 2: Fine-Grained Detection}
\label{detector:phase2}
For the vulnerable smart contracts identified in the first phase, we apply node classification on the node embeddings of their Heterogeneous Contract Graphs
to identify the nodes that may contain vulnerabilities, which correspond to statements or lines of code and
allow us to detect the locations of the vulnerabilities at 
the fine-grained line level
in smart contract source code.

\section{Empirical Evaluation}
\label{sec:eval}

This section presents our experimental settings and results to answer these research questions: 
\textbf{RQ1:} The performance of our models compared to several state-of-the-art baselines on contract-level vulnerability classification,
and \textbf{RQ2:} The performance of our models on line-level vulnerability detection.

\subsection{Datasets}
Our evaluation is carried out on a mixed dataset from three datasets: 
\textbf{(1) Smartbugs Curated}~\cite{SmartBugs1,SmartBugs2} is a collection of vulnerable Ethereum smart contracts organized into nine types.
This dataset is one of the most used real datasets for research in automated reasoning and testing of smart contracts written in Solidity.
It contains 143 annotated contracts having 208 tagged vulnerabilities. 
\textbf{(2) SolidiFI-Benchmark}~\cite{ghaleb2020effective} is a synthetic dataset of vulnerable smart contracts.
There are 9369 injected vulnerabilities in 350 distinct contracts, with seven different vulnerability types.  
To ensure consistency in the evaluation, we only focus on the seven types of vulnerabilities that are joint in both datasets, including: 
    \textit{Access Control},
    \textit{Arithmetic},
    \textit{Denial of Service},
    \textit{Front Running}
    \textit{Reentracy},
    \textit{Time manipulation},
    and \textit{Unchecked Low Level Calls}.
 \textbf{(3) Clean Smart Contracts from Smartbugs Wild}~\cite{SmartBugs1, SmartBugs2} is a collection of 47,398 unique smart contracts from the Ethereum network. Based on the results of eleven integrated detection tools, the Smartbugs framework reports 2,742 contracts that do not contain any bugs, out of the 47,398 contracts. 
Thus, we use the 2,742 contracts as a set of clean contracts. 

\smallskip
For the coarse-grained contract-level vulnerability classification tasks, we randomly take some smart contracts from the clean set and then mix them with the Smartbugs Curated and SolidiFi-Benchmark sets.
We keep a ratio of 1:1 between clean and buggy contracts since this helps us create more balanced train/test sets for the tasks since there are only from 44 to 95 buggy contracts labeled per each bug type (see Table~\ref{tab:dataset_statistics}).
For the fine-grained line-level vulnerability detection tasks, we use the dataset containing vulnerable smart contracts only, i.e., the union of SmartBugs Curated and SolidiFI-Benchmark sets. We do not use other datasets such as the ones of Zhuang \textit{et al.}~\cite{zhuang2020smart}, Liu \textit{et al.}~\cite{liu2021combining} and eThor~\cite{schneidewind2020ethor} because they do not have fine-grained line-level labels for the vulnerabilities.


\smallskip
Note that the Slither parser we use does not automatically generate the clean or vulnerable labels for a node.
Instead, the nodes are labeled based on the lines of vulnerable code either manually by Smartbugs authors or injected by the SolidiFI tool.
For example, Line 15 in Figure~\ref{fig:code-snippet} contains a Reentrancy bug labeled by Smartbugs; then, the nodes with red text in the Heterogeneous CFG and CG are labeled vulnerable.



\subsection{Comparison Methods}

\subsubsection{Comparison to Graph-based neural network Methods} We use the four state-of-the-art methods, including: 
\textit{node2vec}~\cite{node2vec} learns node embeddings by minimizing the cross-entropy loss between the embedding of two nodes belonging to the same random walk with negative sampling;
\textit{LINE}~\cite{tang2015line} only differs from node2vec in the exact formulations of the loss functions and optimizing strategies; 
\textit{Graph Convolutional Network (GCN)}~\cite{GCN} generalizes the convolutional neural network 
by using the Laplacian matrix as a first-order approximation for the propagation among the layers of spectral graph convolutions;
and \textit{metapath2vec}~\cite{metapath2vec} maximizes the likelihood of retaining the structures and semantics of the node/edge labels using the embedding of each node in heterogeneous graphs.
Note that the original architectures of node2vec, LINE, GCN, and metapath2vec only focus on graph topology and do not have any components to handle node features.

Although HAN~\cite{HAN} inspired some idea for our Node-Level Attention Heterogeneous Graph Neural Network, our approach has novelty in resolving the challenges of fitting with the customized metapaths that the original HAN model could not handle effectively. In particular, HAN requires a predefined list of metapaths and each HAN model only serves one or some predefined node types. However, the \ourapp's Heterogeneous CFGs and CGs have dynamic types of nodes and edges, leading to 
difficulties in predefining metapaths like the original HAN model, and thus we did not use HAN as a baseline in our evaluation.


The output embeddings of the homogeneous and heterogeneous graph neural networks are used in two ways in our evaluation. First, we use them directly as the baselines for the coarse-grained graph classification tasks and fine-grained node classification tasks. 
Second, each of the graph neural networks is plugged into \ourapp as the topological graph neural network. The generated embeddings are then considered as the node features fed to \ourapp's Node-Level Attention Heterogeneous Graph Neural Network. Besides, we use fully-connected layers as the multi-layer perceptron in node and graph classification tasks.
In addition, the one-hot vectors based on the Node-Type is also used as the node features, which allows \ourapp to perform independently without relying on any added-in topological graph neural network.

\textbf{Parameter Settings:} The node embedding size is set to 128 for all models. We use adaptive learning rate from 0.0005 to 0.01 in coarse-grained tasks and from 0.0002 to 0.005 in fine-grained tasks when training. 
For each GAT layer~\cite{GAT} of each metapath that feeds to the \ourapp's Self-Attention Layer per Node Type, we set 8 multi-heads whose hidden size is 32. The numbers of learning epochs of coarse-grained and fine-grained tasks are 50 and 100, respectively, to reach converging. 
For node2vec, LINE, GCN, and metapath2vec, we use the authors' recommended settings to ensure the highest performance.

\subsubsection{Comparison with Conventional Detection Tools}

We also compare our method to six common smart contract vulnerability detection tools based on traditional software engineering approaches:
\textit{Manticore}~\cite{mossberg2019manticore} analyzes the symbolic execution of smart contracts and binaries;
\textit{Mythril}~\cite{Mueller2018} uses symbolic execution, SMT solving, and taint analysis to find out the security vulnerabilities of smart contracts;
\textit{Oyente}~\cite{Oyente} analyzes symbolic execution to detect bugs in the Ethereum blockchain;
\textit{Securify}~\cite{tsankov2018securify} can prove if the behavior of a smart contract is safe or not according to given predicates and by checking its graph dependencies;
\textit{Slither}~\cite{Slither} reduces the complexity of instruction sets with the intermediate representation of Ethereum smart contract called SlithIR, while retaining much of the semantic to increase the accuracy of bug detection;
\textit{Smartcheck}~\cite{tikhomirov2018smartcheck} converts smart contracts into XML-based representation and finds possible bugs along executive paths.

\subsection{Evaluation Metrics}

\begin{table*}
\centering
\footnotesize
\resizebox{0.95\textwidth}{!}{
\begin{tabular}{|l|l||l||l|l|l|l|l|l|l|} 
\hline
\multicolumn{2}{|c||}{\textbf{Methods }}                                                                                                                                                                                                                                            & \multicolumn{1}{c||}{\textbf{Metrics}}                         & \multicolumn{1}{c|}{\begin{tabular}[c]{@{}c@{}}\textbf{Access }\\\textbf{Control}\end{tabular}} & \multicolumn{1}{c|}{\textbf{Arithmetic}} & \multicolumn{1}{c|}{\begin{tabular}[c]{@{}c@{}}\textbf{Denial of }\\\textbf{Service}\end{tabular}} & \multicolumn{1}{c|}{\begin{tabular}[c]{@{}c@{}}\textbf{Front }\\\textbf{Running}\end{tabular}} & \multicolumn{1}{c|}{\textbf{Reentrancy}} & \multicolumn{1}{c|}{\begin{tabular}[c]{@{}c@{}}\textbf{Time }\\\textbf{Manipulation}\end{tabular}} & \multicolumn{1}{c|}{\begin{tabular}[c]{@{}c@{}}\textbf{Unchecked Low }\\\textbf{Level Calls}\end{tabular}}  \\ 

\hline \hline
\multirow{2}{*}{\begin{tabular}[c]{@{}l@{}}Heterogeneous GNN\end{tabular}} 
& \multirow{2}{*}{metapath2vec}  
& \begin{tabular}[c]{@{}l@{}}Buggy F1\end{tabular} 
& 62.90\%	& 56.46\%	& 55.17\%	& 63.40\%	& 61.79\%	& 66.29\%	& 55.22\%  \\
\cline{3-10}  &  & \begin{tabular}[c]{@{}l@{}}Macro-F1\end{tabular}
& 42.55\%	& 46.32\%	& 44.49\%	& 43.03\%	& 47.26\%	& 45.94\%	& 49.05\%  \\
\hline \hline
\multirow{6}{*}{\begin{tabular}[c]{@{}l@{}}Homogeneous GNNs\end{tabular}}           & \multirow{2}{*}{GCN}                                                                         & \begin{tabular}[c]{@{}l@{}}Buggy F1\end{tabular} 
& 60.63\%	& -	& 60.12\%	& -	& -	& 59.60\%	& -  \\
\cline{3-10} & & \begin{tabular}[c]{@{}l@{}}Macro-F1\end{tabular}  
& 48.45\%	& -	& 45.65\%	& -	& -	& 46.60\%	& -  \\
\cline{2-10} & \multirow{2}{*}{LINE} & \begin{tabular}[c]{@{}l@{}}Buggy F1\end{tabular} 
& 61.45\%	& 33.41\%	& 59.61\%	& 62.61\%	& 66.23\%	& 66.65\%	& 60.51\%  \\
\cline{3-10} & & \begin{tabular}[c]{@{}l@{}}Macro-F1\end{tabular}     
& 40.88\%	& 33.47\%	& 35.77\%	& 34.29\%	& 37.91\%	& 40.84\%	& 40.08\%  \\
\cline{2-10} & \multirow{2}{*}{node2vec}                                            & \begin{tabular}[c]{@{}l@{}}Buggy F1\end{tabular} 
& 62.63\%	& 58.59\%	& 56.41\%	& 64.77\%	& 58.29\%	& 63.03\%	& 61.69\%  \\
\cline{3-10} &    & \begin{tabular}[c]{@{}l@{}}Macro-F1\end{tabular} 
& 48.83\%	& 50.80\%	& 40.63\%	& 46.08\%	& 45.80\%	& 46.78\%	& 49.91\%  \\

\hline \hline
\multirow{10}{*}{\begin{tabular}[c]{@{}l@{}}\textbf{\ourapp with}\\\textbf{Node Features }\\\textbf{Generated by }\end{tabular}} & \multirow{2}{*}{
\begin{tabular}[c]{@{}l@{}}\textbf{NodeType One}\\\textbf{Hot Vectors}\end{tabular}} & \begin{tabular}[c]{@{}l@{}}Buggy F1\end{tabular} 
& \textbf{71.19\%} 	& \textbf{66.85\%} 	& 87.37\% 	& 87.31\% 	& \textbf{76.09\%} 	& 85.03\% 	& \textbf{72.08\%}  \\
\cline{3-10}  &  & \begin{tabular}[c]{@{}l@{}}Macro-F1 \end{tabular}
& \textbf{74.57\%} 	& \textbf{71.04\%} 	& 86.68\% 	& 85.65\% 	& \textbf{75.80\%} 	& 83.35\% 	& \textbf{74.52\%}  \\
\cline{2-10} & \multirow{2}{*}{\textbf{metapath2vec }}
& \begin{tabular}[c]{@{}l@{}}Buggy F1\end{tabular} 
& 57.70\% 	& 52.84\% 	& 60.16\% 	& 62.19\% 	& 55.06\% 	& 59.47\% 	& 51.37\%  \\
\cline{3-10} &   & \begin{tabular}[c]{@{}l@{}}Macro-F1\end{tabular}
& 55.60\% 	& 55.06\% 	& 64.12\% 	& 64.80\% 	& 60.96\% 	& 57.74\% 	& 55.58\%  \\
\cline{2-10} & \multirow{2}{*}{\textbf{GCN }}
& \begin{tabular}[c]{@{}l@{}}Buggy F1\end{tabular} 
& 49.26\%   & - 	    & 53.19\%   & - 	    & -         & 49.50\%   & -  \\
\cline{3-10} & & \begin{tabular}[c]{@{}l@{}}Macro-F1\end{tabular} 
& 52.75\%   & -         & 60.26\%   & -         & - 	    & 57.31\%   & -  \\
\cline{2-10} & \multirow{2}{*}{\textbf{LINE }} 
& \begin{tabular}[c]{@{}l@{}}Buggy F1\end{tabular} 
& 65.12\% 	& 54.91\% 	& \textbf{89.15\%} 	& \textbf{89.86\%} 	& 71.04\% 	& \textbf{87.71\%} 	& 59.44\%  \\
\cline{3-10} &  & \begin{tabular}[c]{@{}l@{}}Macro-F1\end{tabular} 
& 70.15\% 	& 65.36\% 	& \textbf{89.46\%} 	& \textbf{88.66\%} 	& 74.97\% 	& \textbf{86.41\%} 	& 66.16\%  \\
\cline{2-10} & \multirow{2}{*}{\textbf{node2vec }} & \begin{tabular}[c]{@{}l@{}}Buggy F1\end{tabular} 
& 55.71\% 	& 64.11\% 	& 83.86\% 	& 86.05\% 	& 71.39\% 	& 73.38\% 	& 66.10\%  \\
\cline{3-10} &  & \begin{tabular}[c]{@{}l@{}}Macro-F1\end{tabular}
& 64.70\% 	& 70.23\% 	& 83.40\% 	& 84.95\% 	& 72.31\% 	& 74.36\% 	& 71.02\%  \\
\hline
\end{tabular}}
\caption{\small Average Performance Comparison of the Coarse-Grained Contract-Level Detection over 20 Runs. We use the \textit{Heterogeneous Contract Graphs} of both Clean and Buggy Smart Contracts as the \ourapp framework inputs. \textit{Buggy- F1} means the F1-score of the buggy graph label. ‘–’ denotes not applicable due to the insufficiency of GPU memory to handle the input graphs for the GCN model.} 
\label{tab:f0_cfg}
\end{table*}
Since our prediction results are based on binary classification of a node or a graph, we use F1-score and Macro-F1 scores to measure the prediction performance.
The former is a measure of a model’s performance by balancing between precision and recall,
while the latter is used to assess the quality of problems with multiple binary labels or multiple classes. 
In our evaluation, the F1-score metric is used to evaluate the models' performance when finding vulnerabilities in the graphs, and we also call it {\it Buggy-F1}.
Macro-F1 is considered to avoid biases in the clean and vulnerability labels.

\subsection{Empirical Results}

\begin{table}[tb]
\centering
\footnotesize
\resizebox{0.95\columnwidth}{!}{
\begin{tabular}{l|l|l|l|l}
\textbf{Bug Types}                                                                              & \begin{tabular}[c]{@{}l@{}}\textbf{\# Total / Buggy}\\\textbf{Contracts}\end{tabular} & \begin{tabular}[c]{@{}l@{}}\textbf{\# Total}\\\textbf{Nodes}\end{tabular} & \begin{tabular}[c]{@{}l@{}}\textbf{\# Total}\\\textbf{Edges}\end{tabular} & \begin{tabular}[c]{@{}l@{}}\textbf{\# Buggy}\\\textbf{Nodes}\end{tabular}  \\ 
\hline
\begin{tabular}[c]{@{}l@{}}\textbf{Access }\\\textbf{Control}\end{tabular}                      & 114 / 57                                                                            & 13014                                                                     & 10721                                                                     & 7500                                                                       \\ 
\hline
\textbf{Arithmetic}                                                                             & 120 / 60                                                                            & 17372                                                                     & 14271                                                                     & 10110                                                                      \\ 
\hline
\begin{tabular}[c]{@{}l@{}}\textbf{Denial of }\\\textbf{Service}\end{tabular}                   & 92 / 46                                                                             & 13968                                                                     & 11997                                                                     & 8280                                                                       \\ 
\hline
\begin{tabular}[c]{@{}l@{}}\textbf{Front}\\\textbf{Running}\end{tabular}                        & 88 / 44                                                                             & 22824                                                                     & 19761                                                                     & 10008                                                                      \\ 
\hline
\textbf{Reentrancy}                                                                             & 142 / 71                                                                            & 18898                                                                     & 17614                                                                     & 11238                                                                      \\ 
\hline
\begin{tabular}[c]{@{}l@{}}\textbf{Time~}\\\textbf{Manipulation}\end{tabular}                   & 100 / 50                                                                            & 16765                                                                     & 15550                                                                     & 10051                                                                      \\ 
\hline
\begin{tabular}[c]{@{}l@{}}\textbf{Unchecked }\\\textbf{Low~Level}\\\textbf{Calls}\end{tabular} & 190 / 95                                                                            & 17756                                                                     & 14858                                                                     & 7583                                                            
\end{tabular}}
\caption{\small Statistics of the Mixed Dataset}
\label{tab:dataset_statistics}
\end{table}
Table~\ref{tab:dataset_statistics} shows the statistics of the mixed dataset. In the initial experiments, we split the dataset into 60\%/20\%/20\% for the corresponding train/validation/test sets. However, some bug types in our mixed dataset have less than 100 contracts, which leads to a lack of enough samples for training. Besides, we realized that the loss value remains stable after a fixed number of epochs (100 and 50 epochs for Fine-Grained for Coarse-Grained tasks, respectively). Hence, we decided to split the dataset to 70\%/30\% to increase the train/test set sizes and maintain the vulnerable nodes' ratio in each set corresponding to the whole dataset. To get robust results for each dataset, each embedding method, and each vulnerability type, we run the experiment twenty times independently, each time with a different random seed, and report the average results. Besides, our approach shows impressive capabilities in training and inference time. It takes around 30 seconds for over ten thousand nodes and edges in the node classification task and under 10 seconds for about 100--200 contracts in the graph classification task. Also, it requires under 1 second for all inferences.

\subsubsection{Coarse-Grained Contract-Level Vulnerability Detection
(\textbf{RQ1})}
\label{sec:coarse-grainedresults}
In this experiment, we want to measure \ourapp's performance with various node feature generator components in detecting vulnerable smart contracts (see Section~\ref{detector:phase1}).
It illustrates the flexibility of our method working with different graph neural networks.
Table~\ref{tab:f0_cfg} presents \ourapp's performance via several different graph neural methods on various vulnerability types. Accordingly, we have some observations:
\begin{itemize}[nosep,leftmargin=1em]
    \item \ourapp generally outperforms baseline GNNs in contract-level detection. For instance, the Buggy-F1 and Macro-F1 of \ourapp are over 88.66\%, while the maximum performance of the baselines is 64.77\% in detecting the Front-Running vulnerability type.
    \item It is unclear which node feature generation method is the best among the heterogeneous and homogeneous GNNs and the node-type one-hot vectors. 
    However, integrating these types of GNNs inside \ourapp outperforms all the baselines. Hence, we believe that the architecture of \ourapp for combining different GNNs is suitable for classifying vulnerable smart contracts.
    \item \ourapp is reliable in determining whether an unknown smart contract contains vulnerabilities, especially for the vulnerability types of Denial of Service, Front Running, and Time Manipulation with Buggy-F1 over 87.7\%. 
    \ourapp is highly compatible with different solidity versions based on the Slither tool \cite{Slither}, and its trained models can be applied in practice to audit newly-appeared smart contracts that previous studies using graph learning~\cite{zhuang2020smart, liu2021smart} have not been able to do effectively (see Section~\ref{subsec:related-works-ge-nn}).
\end{itemize}

\subsubsection{Fine-Grained Line-Level Vulnerability Detection (\textbf{RQ2})}
\label{sec:fine-grainedresults} 

\begin{table*}
\centering
\footnotesize
\resizebox{0.92\textwidth}{!}{
\begin{tabular}{|l|l||l||l|l|l|l|l|l|l|} 
\hline
\multicolumn{2}{|c||}{\textbf{Methods }}                                                                                                                                                                                                                                            & \multicolumn{1}{c||}{\textbf{Metrics}}                         & \multicolumn{1}{c|}{\begin{tabular}[c]{@{}c@{}}\textbf{Access }\\\textbf{Control}\end{tabular}} & \multicolumn{1}{c|}{\textbf{Arithmetic}} & \multicolumn{1}{c|}{\begin{tabular}[c]{@{}c@{}}\textbf{Denial of }\\\textbf{Service}\end{tabular}} & \multicolumn{1}{c|}{\begin{tabular}[c]{@{}c@{}}\textbf{Front }\\\textbf{Running}\end{tabular}} & \multicolumn{1}{c|}{\textbf{Reentrancy}} & \multicolumn{1}{c|}{\begin{tabular}[c]{@{}c@{}}\textbf{Time }\\\textbf{Manipulation}\end{tabular}} & \multicolumn{1}{c|}{\begin{tabular}[c]{@{}c@{}}\textbf{Unchecked Low }\\\textbf{Level Calls}\end{tabular}}  \\ 
\hline \hline
\multirow{12}{*}{\begin{tabular}[c]{@{}l@{}}Conventional \\Detection Tools\end{tabular}}                                                                                    & \multirow{2}{*}{securify}                                                                            & \begin{tabular}[c]{@{}l@{}}Buggy F1\end{tabular} & 13.0\%                                                                                          & 0.0\%                                    & 18.0\%                                                                                             & 53.0\%                                                                                         & 23.0\%                                   & 24.0\%                                                                                             & 11.0\%                                                                                                      \\ 
\cline{3-10}
                                                                                                                                                                            &                                                                                                      & \begin{tabular}[c]{@{}l@{}}Macro-F1\end{tabular}     & 52.3\%                                                                                          & 45.2\%                                   & 52.0\%                                                                                             & 72.2\%                                                                                         & 58.4\%                                   & 52.4\%                                                                                             & 54.1\%                                                                                                      \\ 
\cline{2-10}
                                                                                                                                                                            & \multirow{2}{*}{mythril}                                                                             & \begin{tabular}[c]{@{}l@{}}Buggy F1\end{tabular} & 34.0\%                                                                                          & 73.0\%                                   & 41.0\%                                                                                             & 63.0\%                                                                                         & 19.0\%                                   & 23.0\%                                                                                             & 14.0\%                                                                                                      \\ 
\cline{3-10}
                                                                                                                                                                            &                                                                                                      & \begin{tabular}[c]{@{}l@{}}Macro-F1\end{tabular}     & 61.1\%                                                                                          & \textbf{84.1\%}                          & 60.1\%                                                                                             & 77.8\%                                                                                         & 55.3\%                                   & 50.8\%                                                                                             & 55.7\%                                                                                                      \\ 
\cline{2-10}
                                                                                                                                                                            & \multirow{2}{*}{slither}                                                                             & \begin{tabular}[c]{@{}l@{}}Buggy F1\end{tabular} & 32.0\%                                                                                          & 0.0\%                                    & 13.0\%                                                                                             & 26.0\%                                                                                         & 15.0\%                                   & 44.0\%                                                                                             & 10.0\%                                                                                                      \\ 
\cline{3-10}
                                                                                                                                                                            &                                                                                                      & \begin{tabular}[c]{@{}l@{}}Macro-F1\end{tabular}     & 61.5\%                                                                                          & 45.2\%                                   & 42.7\%                                                                                             & 56.9\%                                                                                         & 49.4\%                                   & 57.3\%                                                                                             & 53.3\%                                                                                                      \\ 
\cline{2-10}
                                                                                                                                                                            & \multirow{2}{*}{manticore}                                                                           & \begin{tabular}[c]{@{}l@{}}Buggy F1\end{tabular} & 30.0\%                                                                                          & 30.0\%                                   & 12.0\%                                                                                             & 7.0\%                                                                                          & 9.0\%                                    & 24.0\%                                                                                             & 4.0\%                                                                                                       \\ 
\cline{3-10}
                                                                                                                                                                            &                                                                                                      & \begin{tabular}[c]{@{}l@{}}Macro-F1\end{tabular}     & 61.1\%                                                                                          & 61.0\%                                   & 48.0\%                                                                                             & 46.9\%                                                                                         & 51.2\%                                   & 55.1\%                                                                                             & 50.6\%                                                                                                      \\ 
\cline{2-10}
                                                                                                                                                                            & \multirow{2}{*}{smartcheck}                                                                          & \begin{tabular}[c]{@{}l@{}}Buggy F1\end{tabular} & 20.0\%                                                                                          & 22.0\%                                   & 52.0\%                                                                                             & 0.0\%                                                                                          & 22.0\%                                   & 44.0\%                                                                                             & 11.0\%                                                                                                      \\ 
\cline{3-10}
                                                                                                                                                                            &                                                                                                      & \begin{tabular}[c]{@{}l@{}}Macro-F1\end{tabular}     & 56.0\%                                                                                          & 56.1\%                                   & 69.9\%                                                                                             & 46.2\%                                                                                         & 57.8\%                                   & 64.2\%                                                                                             & 54.1\%                                                                                                      \\ 
\cline{2-10}
                                                                                                                                                                            & \multirow{2}{*}{oyente}                                                                              & \begin{tabular}[c]{@{}l@{}}Buggy F1\end{tabular} & 21.0\%                                                                                          & 71.0\%                                   & 48.0\%                                                                                             & 0.0\%                                                                                          & 20.0\%                                   & 24.0\%                                                                                             & 8.0\%                                                                                                       \\ 
\cline{3-10}
                                                                                                                                                                            &                                                                                                      & \begin{tabular}[c]{@{}l@{}}Macro-F1\end{tabular}     & 57.3\%                                                                                          & 82.8\%                                   & 67.2\%                                                                                             & 44.8\%                                                                                         & 56.1\%                                   & 52.4\%                                                                                             & 52.6\%                                                                                                      \\ 
\hline \hline
\multirow{2}{*}{\begin{tabular}[c]{@{}l@{}}Heterogeneous GNN\end{tabular}}                                                                       & \multirow{2}{*}{metapath2vec}                                                                        & \begin{tabular}[c]{@{}l@{}}Buggy F1\end{tabular} &  35.46\%  & 	68.70\%  & 	60.64\%  & 	80.65\%  & 	71.66\%  & 	67.51\%   &	26.06\%  \\
\cline{3-10}  &   & \begin{tabular}[c]{@{}l@{}}Macro-F1\end{tabular} 
&  48.52\%  & 	47.08\%  & 	48.67\%  & 	49.88\%  & 	49.15\%  & 	49.00\%   &	49.91\%  \\
\hline \hline
\multirow{6}{*}{\begin{tabular}[c]{@{}l@{}}Homogeneous GNNs\end{tabular}}    & \multirow{2}{*}{GCN} & \begin{tabular}[c]{@{}l@{}}Buggy F1\end{tabular} 
&  43.92\%  & 	65.69\%  & 	64.06\%  & 	81.09\%  & 	71.76\%  & 	68.70\%   &	38.13\%  \\
\cline{3-10} &    & \begin{tabular}[c]{@{}l@{}}Macro-F1\end{tabular}
&  54.20\%  & 	53.42\%  & 	54.81\%  & 	56.21\%  & 	53.00\%  & 	52.74\%   &	53.57\%  \\
\cline{2-10} & \multirow{2}{*}{LINE}  & \begin{tabular}[c]{@{}l@{}}Buggy F1\end{tabular} 
&  53.59\%  & 	68.61\%  & 	62.28\%  & 	83.06\%  & 	74.78\%  & 	70.76\%  & 	7.10\%  \\
\cline{3-10} &     & \begin{tabular}[c]{@{}l@{}}Macro-F1\end{tabular}
&  57.75\%  & 	48.53\%  & 	51.63\%  & 	42.27\%  & 	38.26\%  & 	42.40\%   &	44.31\%  \\
\cline{2-10}  & \multirow{2}{*}{node2vec}   & \begin{tabular}[c]{@{}l@{}}Buggy F1\end{tabular} 
&  44.94\%  & 	67.84\%  & 	63.92\%  & 	81.84\%  & 	71.52\%  & 	67.81\%   &	34.26\%  \\
\cline{3-10}  &    & \begin{tabular}[c]{@{}l@{}}Macro-F1\end{tabular} 
&  54.73\%  & 	52.92\%  & 	54.83\%  & 	56.17\%  & 	53.45\%  & 	53.19\%   &	53.09\%  \\
\hline \hline
\multirow{10}{*}{\begin{tabular}[c]{@{}l@{}}\textbf{\ourapp with}\\\textbf{Node Features }\\\textbf{Generated by }\end{tabular}} & \multirow{2}{*}{\begin{tabular}[c]{@{}l@{}}\textbf{NodeType One}\\\textbf{Hot Vectors}\end{tabular}} & \begin{tabular}[c]{@{}l@{}}Buggy F1\end{tabular} 
& 77.21\%	& 81.62\%	& 79.83\%	& 88.19\%	& 84.24\%	& 86.64\% &	65.95\%  \\
\cline{3-10}  &  & \begin{tabular}[c]{@{}l@{}}Macro-F1 \end{tabular}
& 74.89\%	& 76.01\%	& 76.22\%	& 68.70\%	& 75.89\%	& 82.72\% &	75.01\%  \\
\cline{2-10} & \multirow{2}{*}{\textbf{metapath2vec }}
& \begin{tabular}[c]{@{}l@{}}Buggy F1\end{tabular} 
& 67.97\%	& 74.84\%	& 67.22\%	& 86.08\%	& 76.03\%	& 73.81\% &	50.71\%  \\
\cline{3-10} &   & \begin{tabular}[c]{@{}l@{}}Macro-F1\end{tabular}
& 67.87\%	& 65.92\%	& 62.90\%	& 65.22\%	& 66.04\%	& 71.04\% &	64.73\%  \\
\cline{2-10} & \multirow{2}{*}{\textbf{GCN }}
& \begin{tabular}[c]{@{}l@{}}Buggy F1\end{tabular} 
& 69.00\%	& 76.47\%	& 70.88\%	& 87.15\%	& 77.57\%	& 77.73\% &	52.95\%  \\
\cline{3-10} & & \begin{tabular}[c]{@{}l@{}}Macro-F1\end{tabular} 
& 66.77\%	& 66.75\%	& 64.26\%	& 65.71\%	& 65.85\%	& 73.94\% &	65.75\%  \\
\cline{2-10} & \multirow{2}{*}{\textbf{LINE }} 
& \begin{tabular}[c]{@{}l@{}}Buggy F1\end{tabular} 
& 81.19\%	& 81.58\%	& \textbf{82.12\%}	& 90.47\%	& 86.27\%	& 89.21\% &	83.37\%  \\
\cline{3-10} &  & \begin{tabular}[c]{@{}l@{}}Macro-F1\end{tabular} 
& \textbf{80.93\%}	& 77.80\%	& \textbf{79.00\%}	& 78.43\%	& 80.43\%	& 86.17\% &	85.40\%  \\
\cline{2-10} & \multirow{2}{*}{\textbf{node2vec }} & \begin{tabular}[c]{@{}l@{}}Buggy F1\end{tabular} 
& \textbf{81.98\%}	& \textbf{84.35\%}	& 82.09\%	& \textbf{90.51\%}	& \textbf{86.40\%}	& \textbf{90.29\%} &	\textbf{84.81\%}  \\
\cline{3-10} &  & \begin{tabular}[c]{@{}l@{}}Macro-F1\end{tabular}
& 79.23\%	& 79.10\%	& 77.84\%	& \textbf{78.60\%}	& \textbf{80.78\%}	& \textbf{86.76\%} &	\textbf{86.74\%}  \\
\hline
\end{tabular}}
\caption{\small Average Performance Comparison of the Fine-Grained Line-Level Detection over 20 Runs. We use the  \textit{Heterogeneous Contract Graphs} of the Buggy Smart Contracts as the inputs for \ourapp framework. \textit{Buggy- F1} means the F1-score of the buggy node label.
A total of fifteen methods are examined in the comparisons. The best performance in each vulnerability category is highlighted.}
\label{tab:f1_cfg}
\end{table*}
To help smart contract developers to locate vulnerabilities more easily, vulnerability detectors should be able to identify the vulnerabilities at the more fine-grained line level (see Section~\ref{detector:phase2}).
In this experiment, we examine the performance of our method with respect to various state-of-the-art methods for line-level detection.

Table~\ref{tab:f1_cfg} shows the performance of our method trained with different models for Topological Graph Neural Network and the baselines methods, including graph-based neural networks and the conventional detection tools based on various software engineering techniques.
From the table, we observe:
\begin{itemize}[nosep,leftmargin=1em]
    \item Generally, \ourapp outperforms conventional detection tools significantly. Remarkably, an improvement is up to 63.4\% of \ourapp compared to the best performance of the tools in detecting Reentrancy bugs. We argue the significant improvement is from two sources: First, our constructed heterogeneous graphs retain more CFGs' aspects than other analysis tools. Secondly, our node-level attention module is flexible enough for GNNs to learn the exact locations of vulnerabilities within contracts.
    \item Our method beats the results of the baseline GNNs. Remarkably, the macro-F1 scores of the baseline GNNs are up to 60.5\%, while our models can reach up to 80.78\%. Hence, it is evident modeling the smart contracts as Heterogeneous Contract Graphs can benefit vulnerability prediction.
    \item Conventional detection tools perform well in detecting arithmetic bugs. The phenomenon is reasonable since these tools mostly use symbolic execution and such technique is suitable for detecting arithmetic bugs~\cite{SurveySymExec-CSUR18}. However, \ourapp performance is still on par with the tools and our future work will improve the graph models to learn arithmetic operations better. Besides, some conventional detection tools in Table~\ref{tab:f1_cfg} barely work (with Buggy-F1=0\%) for some vulnerability types due to their intrinsic limits in relying on predefined expert patterns that could not capture these vulnerabilities. 
\end{itemize}

\noindent
\textbf{Expanded Experiments.} We also ran the experiments in Tables~\ref{tab:f0_cfg} and \ref{tab:f1_cfg} with only Heterogeneous CFGs and CGs separately. Overall, these results are worse than the fusion form in the heterogeneous contract graphs reported in the paper. The expanded experiments can be found in our Git repository link.

\section{Related Works}
\label{sec:related}



\subsection{Graph Embedding Neural Networks}
\label{subsec:related-works-ge-nn}
A few studies have detected smart contract vulnerabilities using neural network-based embedding techniques. Zhuang \textit{et al.}~\cite{zhuang2020smart} represent each function's syntactic and semantic structures in smart contracts as a contract graph and propose a degree-free graph convolutional neural network with expert patterns to learn the normalized graphs for vulnerability detection. They also provide more interpretable weights by extracting vulnerability-specific expert patterns for encoding graphs \cite{liu2021smart}.
In their Peculiar tool~\cite{wupeculiar2021issre}, Wu \textit{et al.} present a pretraining technique based on customized data flow graphs of smart contract functions to identify reentrance vulnerabilities. However, their methods face various limitations: Relying on expert patterns, their graph generator only works with some pre-defined Major and Secondary functions before generating the contract graphs, leading to poor performance in the graph generation process compared to MANDO.  
Besides, pre-defined patterns also restrict them to detect only two specified bugs, Reentrancy and Time Manipulation, in Solidity source code. In contrast, the heterogeneous graph structure allows \ourapp to be more general and flexible in exploring different vulnerability types without requiring any pre-definitions. 

Other studies use other forms of embeddings: Zhao \textit{et al.}~\cite{Zhao2021} use word embedding together with similarity detection and Generative Adversarial Networks (GAN) to detect reentrance vulnerabilities dynamically. 
SmartConDetect \cite{jeon2021smartcondetect} treats code fragments as unique sequences of tokens and uses a pre-trained BERT model to identify vulnerable patterns.
SmartEmbed \cite{gao2020checking} employs serialized structured syntax trees to train word2vec and fastText models to recognize vulnerabilities. 
Different from such existing techniques, 
our unique graph encodings can accurately capture the vulnerability patterns and locate fine-grained vulnerabilities at the line level. 

\subsection{Code Representation and Learning}
Software programs have explored learning from heterogeneous graphs for vulnerability detection, code search, and other tasks.   
For example, VulDeePecker \cite{li2018vuldeepecker} uses both syntax structures and dependency slices to represent programs and employ commonly used neural network models to learn the programs' embedding and identify vulnerability patterns for C/C++ programs. 
VulDeeLocator \cite{li2021vuldeelocator} extends the work by adding attention-based granularity refinement to identify fine-grained line-level vulnerability locations. BGNN4VD \cite{cao2021bgnn4vd} also uses combined code representations in the abstract syntax trees and control- and data-flow graphs to learn vulnerability patterns via bilateral graph neural networks for C/C++ programs. 
However, no such study has been done for Solidity smart contracts and our study is the first one. 

\subsection{Bug Detection \& Smart Contracts}

Several studies detect specific types of bugs or vulnerabilities using traditional program analysis and software engineering and security techniques. For example, OYENTE \cite{Oyente} uses symbolic execution to explore execution paths in smart contracts as much as possible and search for four types of bugs.
SmartCheck \cite{tikhomirov2018smartcheck} uses static analysis techniques to check smart contract code for patterns that match pre-defined rules about vulnerabilities and code smells. Several other studies use formal verification to check smart contracts' safety and functional correctness according to certain human-defined specifications~\cite{garfatta2021survey}. 
In addition, many studies are based on abstract interpretation, fuzz testing, enhanced compilation, dynamic consistency checking, and other techniques \cite{chen2020survey,wang2021security}. 
However, in contrast to our automatic bug pattern detection method, such security analysis techniques are built to discover specific vulnerabilities according to manually defined patterns or specifications, limiting their scalability and accuracy.

\section{Conclusion and Future Work}
\label{sec:con}
The popularity and importance of smart contracts in blockchain platforms are increasing. Therefore, it is highly desirable to ensure the quality and security of smart contract programs. 
In this paper, we proposed a new method, based on multi-level graph embeddings of control-flow graphs and call graphs of Solidity smart contracts, to train more accurate vulnerability detection models that can identify vulnerabilities in smart contracts at fine-grained line level and contract level of granularity.
Our evaluation of a large-scale dataset curated from real-world Solidity smart contracts shows that our method is promising and outperforms several baselines. 
Our method is thus a valuable complement to other vulnerability detection techniques and contributes to smart contract security. However, with all the achievements, our method and evaluation can still be improved further. The embedding techniques can further fuse more semantic properties of the smart contract source code, such as data dependencies, and adapt newer and more sophisticated graph neural networks. 
We can also adapt our method to cases where only compiled smart contract bytecode is available without source code to expand. 
The evaluation can further compare with vulnerability detection techniques developed for other programming languages (e.g., C/C++, Java) to check the generalizability of our method.


\balance
\bibliographystyle{IEEEtran}
\bibliography{references}

\begin{thebibliography}{10}
\providecommand{\url}[1]{#1}
\csname url@samestyle\endcsname
\providecommand{\newblock}{\relax}
\providecommand{\bibinfo}[2]{#2}
\providecommand{\BIBentrySTDinterwordspacing}{\spaceskip=0pt\relax}
\providecommand{\BIBentryALTinterwordstretchfactor}{4}
\providecommand{\BIBentryALTinterwordspacing}{\spaceskip=\fontdimen2\font plus
\BIBentryALTinterwordstretchfactor\fontdimen3\font minus
  \fontdimen4\font\relax}
\providecommand{\BIBforeignlanguage}[2]{{%
\expandafter\ifx\csname l@#1\endcsname\relax
\typeout{** WARNING: IEEEtran.bst: No hyphenation pattern has been}%
\typeout{** loaded for the language `#1'. Using the pattern for}%
\typeout{** the default language instead.}%
\else
\language=\csname l@#1\endcsname
\fi
#2}}
\providecommand{\BIBdecl}{\relax}
\BIBdecl

\bibitem{HAN}
X.~Wang, H.~Ji, C.~Shi, B.~Wang, Y.~Ye, P.~Cui, and P.~S. Yu, ``Heterogeneous
  graph attention network,'' in \emph{The World Wide Web Conference}, 2019, pp.
  2022--2032.

\bibitem{wood2014ethereum}
G.~Wood \emph{et~al.}, ``Ethereum: A secure decentralised generalised
  transaction ledger,'' \emph{Ethereum project yellow paper}, vol. 151, no.
  2014, pp. 1--32, 2014.

\bibitem{node2vec}
A.~Grover and J.~Leskovec, ``node2vec: Scalable feature learning for
  networks,'' in \emph{the 22nd ACM SIGKDD International Conference on
  Knowledge Discovery and Data Mining}, 2016, pp. 855--864.

\bibitem{tang2015line}
J.~Tang, M.~Qu, M.~Wang, M.~Zhang, J.~Yan, and Q.~Mei, ``Line: Large-scale
  information network embedding,'' in \emph{WWW}, 2015.

\bibitem{GCN}
T.~N. Kipf and M.~Welling, ``Semi-supervised classification with graph
  convolutional networks,'' \emph{arXiv preprint arXiv:1609.02907}, 2016.

\bibitem{metapath2vec}
Y.~Dong, N.~V. Chawla, and A.~Swami, ``metapath2vec: Scalable representation
  learning for heterogeneous networks,'' in \emph{the 23rd ACM SIGKDD
  International Conference on Knowledge Discovery and Data Mining}, 2017, pp.
  135--144.

\bibitem{tsankov2018securify}
P.~Tsankov, A.~Dan, D.~D. Cohen, A.~Gervais, F.~Buenzli, and M.~Vechev,
  ``Securify: Practical security analysis of smart contracts,'' in \emph{25th
  ACM Conference on Computer and Communications Security}, 2018.

\bibitem{Mueller2018}
B.~Mueller, ``Smashing smart contracts for fun and real profit,'' in \emph{9th
  annual HITB Security Conference}, pp. 2--51.

\bibitem{Slither}
J.~Feist, G.~Grieco, and A.~Groce, ``Slither: a static analysis framework for
  smart contracts,'' in \emph{IEEE/ACM 2nd International Workshop on Emerging
  Trends in Software Engineering for Blockchain}, 2019, pp. 8--15.

\bibitem{mossberg2019manticore}
M.~Mossberg, F.~Manzano, E.~Hennenfent, A.~Groce, G.~Grieco, J.~Feist,
  T.~Brunson, and A.~Dinaburg, ``Manticore: A user-friendly symbolic execution
  framework for binaries and smart contracts,'' in \emph{the 34th IEEE/ACM
  International Conference on Automated Software Engineering}, 2019, pp.
  1186--1189.

\bibitem{tikhomirov2018smartcheck}
S.~Tikhomirov, E.~Voskresenskaya, I.~Ivanitskiy, R.~Takhaviev, E.~Marchenko,
  and Y.~Alexandrov, ``{SmartCheck}: Static analysis of ethereum smart
  contracts,'' in \emph{the 1st International Workshop on Emerging Trends in
  Software Engineering for Blockchain}, 2018, pp. 9--16.

\bibitem{Oyente}
L.~Luu, D.-H. Chu, H.~Olickel, P.~Saxena, and A.~Hobor, ``Making smart
  contracts smarter,'' in \emph{the ACM SIGSAC conference on computer and
  communications security}, 2016, pp. 254--269.

\bibitem{zhuang2020smart}
Y.~Zhuang, Z.~Liu, P.~Qian, Q.~Liu, X.~Wang, and Q.~He, ``Smart contract
  vulnerability detection using graph neural network.'' in \emph{IJCAI}, 2020,
  pp. 3283--3290.

\bibitem{liu2021smart}
Z.~Liu, P.~Qian, X.~Wang, L.~Zhu, Q.~He, and S.~Ji, ``Smart contract
  vulnerability detection: From pure neural network to interpretable graph
  feature and expert pattern fusion,'' \emph{arXiv preprint arXiv:2106.09282},
  2021.

\bibitem{chen2020survey}
H.~Chen, M.~Pendleton, L.~Njilla, and S.~Xu, ``A survey on ethereum systems
  security: Vulnerabilities, attacks, and defenses,'' \emph{ACM Computing
  Surveys (CSUR)}, vol.~53, no.~3, pp. 1--43, 2020.

\bibitem{sun2011pathsim}
Y.~Sun, J.~Han, X.~Yan, P.~S. Yu, and T.~Wu, ``Pathsim: Meta path-based top-k
  similarity search in heterogeneous information networks,'' \emph{the VLDB
  Endowment}, vol.~4, no.~11, pp. 992--1003, 2011.

\bibitem{transformer}
A.~Vaswani, N.~Shazeer, N.~Parmar, J.~Uszkoreit, L.~Jones, A.~N. Gomez,
  {\L}.~Kaiser, and I.~Polosukhin, ``Attention is all you need,'' in
  \emph{Advances in neural information processing systems}, 2017, pp.
  5998--6008.

\bibitem{DeepLearning}
Y.~LeCun, Y.~Bengio, and G.~Hinton, ``Deep learning,'' \emph{nature}, vol. 521,
  no. 7553, pp. 436--444, 2015.

\bibitem{liu2021combining}
Z.~Liu, P.~Qian, X.~Wang, Y.~Zhuang, L.~Qiu, and X.~Wang, ``Combining graph
  neural networks with expert knowledge for smart contract vulnerability
  detection,'' \emph{IEEE Transactions on Knowledge and Data Engineering},
  2021.

\bibitem{SmartBugs1}
T.~Durieux, J.~F. Ferreira, R.~Abreu, and P.~Cruz, ``Empirical review of
  automated analysis tools on 47,587 ethereum smart contracts,'' in \emph{the
  ACM/IEEE 42nd International Conference on Software Engineering}, 2020, pp.
  530--541.

\bibitem{SmartBugs2}
J.~F. Ferreira, P.~Cruz, T.~Durieux, and R.~Abreu, ``Smartbugs: a framework to
  analyze solidity smart contracts,'' in \emph{the 35th IEEE/ACM International
  Conference on Automated Software Engineering}, 2020, pp. 1349--1352.

\bibitem{ghaleb2020effective}
A.~Ghaleb and K.~Pattabiraman, ``How effective are smart contract analysis
  tools? evaluating smart contract static analysis tools using bug injection,''
  in \emph{the 29th ACM SIGSOFT International Symposium on Software Testing and
  Analysis}, 2020.

\bibitem{schneidewind2020ethor}
C.~Schneidewind, I.~Grishchenko, M.~Scherer, and M.~Maffei, ``ethor: Practical
  and provably sound static analysis of ethereum smart contracts,'' in
  \emph{the 2020 ACM SIGSAC Conference on Computer and Communications
  Security}, 2020, pp. 621--640.

\bibitem{GAT}
P.~Veli{\v{c}}kovi{\'c}, G.~Cucurull, A.~Casanova, A.~Romero, P.~Li{\`o}, and
  Y.~Bengio, ``Graph attention networks,'' in \emph{International Conference on
  Learning Representations}, 2018.

\bibitem{SurveySymExec-CSUR18}
R.~Baldoni, E.~Coppa, D.~C. D'Elia, C.~Demetrescu, and I.~Finocchi, ``A survey
  of symbolic execution techniques,'' \emph{ACM Comput. Surv.}, vol.~51, no.~3,
  2018.

\bibitem{wupeculiar2021issre}
H.~Wu, Z.~Zhang, S.~Wang, Y.~Lei, B.~Lin, Y.~Qin, H.~Zhang, and X.~Mao,
  ``Peculiar: Smart contract vulnerability detection based on crucial data flow
  graph and pre-training techniques,'' in \emph{the 32nd International
  Symposium on Software Reliability Engineering}, 2021.

\bibitem{Zhao2021}
H.~Zhao, P.~Su, Y.~Wei, K.~Gai, and M.~Qiu, ``Gan-enabled code embedding for
  reentrant vulnerabilities detection,'' in \emph{Knowledge Science,
  Engineering and Management}, 2021, pp. 585--597.

\bibitem{jeon2021smartcondetect}
S.~Jeon, G.~Lee, H.~Kim, and S.~S. Woo, ``Smartcondetect: Highly accurate smart
  contract code vulnerability detection mechanism using bert,'' in \emph{KDD
  Workshop on Programming Language Processing}, 2021.

\bibitem{gao2020checking}
Z.~Gao, L.~Jiang, X.~Xia, D.~Lo, and J.~Grundy, ``Checking smart contracts with
  structural code embedding,'' \emph{IEEE Transactions on Software
  Engineering}, 2020.

\bibitem{li2018vuldeepecker}
Z.~Li, D.~Zou, S.~Xu, X.~Ou, H.~Jin, S.~Wang, Z.~Deng, and Y.~Zhong,
  ``{VulDeePecker}: A deep learning-based system for vulnerability detection,''
  in \emph{The Network and Distributed System Security Symposium}, 2018.

\bibitem{li2021vuldeelocator}
Z.~Li, D.~Zou, S.~Xu, Z.~Chen, Y.~Zhu, and H.~Jin, ``{VulDeeLocator}: a deep
  learning-based fine-grained vulnerability detector,'' \emph{IEEE Transactions
  on Dependable and Secure Computing}, 2021.

\bibitem{cao2021bgnn4vd}
S.~Cao, X.~Sun, L.~Bo, Y.~Wei, and B.~Li, ``Bgnn4vd: Constructing bidirectional
  graph neural-network for vulnerability detection,'' \emph{Information and
  Software Technology}, vol. 136, p. 106576, 2021.

\bibitem{garfatta2021survey}
I.~Garfatta, K.~Klai, W.~Gaaloul, and M.~Graiet, ``A survey on formal
  verification for solidity smart contracts,'' in \emph{2021 Australasian
  Computer Science Week Multiconference}, 2021, pp. 1--10.

\bibitem{wang2021security}
Y.~Wang, J.~He, N.~Zhu, Y.~Yi, Q.~Zhang, H.~Song, and R.~Xue, ``Security
  enhancement technologies for smart contracts in the blockchain: A survey,''
  \emph{Transactions on Emerging Telecommunications Technologies}, 2021.

\end{thebibliography}

\end{document}